\documentclass[manuscript]{aastex}

\usepackage{color} 
\shorttitle{Fluctuation of the near-infrared background}
\shortauthors{Seo et al.}

\begin{document}

\title{\textit{AKARI} OBSERVATION OF THE SUB-DEGREE SCALE FLUCTUATION OF THE NEAR-INFRARED BACKGROUND}

\author{H. J. Seo\altaffilmark{1}, Hyung Mok Lee\altaffilmark{1}, T. Matsumoto\altaffilmark{2,3,4}, W. -S. Jeong\altaffilmark{4,5}, Myung Gyoon Lee\altaffilmark{1}, and J. Pyo\altaffilmark{4}}

\altaffiltext{1}{Department of Physics and Astronomy, Seoul National University, Seoul 151-742, Korea}
\altaffiltext{2}{Institute of Astronomy and Astrophysics, Academia Sinica, Taipei 10617, Taiwan}
\altaffiltext{3}{Department of Infrared Astrophysics, Institute of Space and Astronautical Science (ISAS), Japan Aerospace Exploration Agency (JAXA), Sagamihara, Kanagawa 252-5210, Japan}
\altaffiltext{4}{Korea Astronomy and Space Science Institute (KASI), Daejeon 305-348, Korea}
\altaffiltext{5}{Korea University of Science and Technology, Daejeon 305-350, Korea}

\email{hjseo@astro.snu.ac.kr}

\begin{abstract}
We report spatial fluctuation analysis of the sky brightness in near-infrared from observations toward the north ecliptic pole (NEP) by the
\textit{AKARI} at 2.4 and 3.2 $\mu$m. As a follow up study of our previous work on the Monitor field of \textit{AKARI}, we used NEP deep survey data, which covered a circular area of about 0.4 square degrees, in order to extend fluctuation analysis at angular scales up to 1000$''$. We found residual fluctuation over the estimated shot noise at larger angles than the angular scale of the Monitor field. The excess fluctuation of the NEP deep field smoothly connects with that of the Monitor field at angular scales with a few hundreds arcseconds and extends without any significant variation to larger angular scales up to 1000$''$. By comparing excess fluctuations at two wavelengths, we confirm a blue spectrum feature similar to the result of the Monitor field. We find that the result of this study is consistent with \textit{Spitzer Space Telescope} observations at 3.6 $\mu$m. The origin of the excess fluctuation in the near-infrared background remains to be answered, but we could exclude zodiacal light, diffuse Galactic light, and unresolved faint galaxies at low-redshift based on the comparison with mid- and far-infrared brightness, ground based near-infrared images.
\end{abstract}


\section{INTRODUCTION}  

It has been known that there exists excess extragalactic background light (EBL) in near-infrared from the data of Diffuse Infrared Background Experiment (DIRBE) on board \textit{Cosmic Background Explorer} (\textit{COBE}; \citealt{hau98,dwe98,wri00,cam01,are03,lev07}) and the Near-infrared Spectrometer (NIRS) on board \textit{Infrared Telescope in Space} (\textit{IRTS}; \citealt{mat05}). Recent analysis of data obtained by \textit{AKARI} \citep{tsu13} also shows a consistent result with them. The excess EBL is several times brighter than the integrated light of galaxies and shows a blue stellar-like spectrum. Since the near-infrared sky brightness is dominated by the zodiacal light, which is scattered sunlight by the interplanetary dust (IPD), the absolute level of the excess EBL has uncertainty due to the inaccuracies in the model of zodiacal light. On the other hand, the fluctuation of near-infrared sky brightness could provide complementary information on the origin of excess emission since the zodiacal light is fairly smooth \citep{pyo12}. Recent studies of \citet{kash05,kash07} and \citet[hereafter Paper 1]{mat11} using the data taken by space telescopes found excess fluctuations of the sky brightness at angular scales of a few hundreds arcseconds that cannot be explained with known emission components. More recently, \citet{coo12b} and \citet{kash12} found an almost flat angular spectrum of the fluctuation extends up to degree angular scales by using data set from \textit{Spitzer} at 3.6 and 4.5 $\mu$m. Characteristic features of observed fluctuation are a blue stellar-like spectrum and a good spatial correlation between different wavelength bands. \citet{zem14} measured significant excess fluctuations at 1.1 and 1.6 $\mu$m using the data from Cosmic Infrared Background Experiment (CIBER).

\defcitealias{mat11}{Paper 1}

One of the intriguing interpretations regarding the origin of the near-infrared excess EBL is the Population III stars (\citealt{kash05}; \citetalias{mat11}) that cannot be identified individually because of their faintness. However, there are some questions and debates about such interpretation. According to \citet{mad05}, large excess of the EBL at near-infrared wavelength bands requires a large number of Population III stars in the early universe and produces too much heavy elements at high redshifts. Also TeV $\gamma$-ray observations prefer lower near-infrared EBL than observed one to explain absorption features in $\gamma$-ray spectra of blazars \citep{aha06,gil12,Meyer12,hess13}. In addition, recent simulations predicted smaller fluctuations than observational measurements by about an order of magnitude \citep{coo12a,fern12,yue13a}. \citet{coo12b} have argued that the intrahalo light (IHL) due to the stars that escaped from galaxies can explain the observed excess fluctuation at large angular scales.

Clearly, more observational studies are necessary to understand the nature of the infrared background radiation. In this study, we extend the scope of our previous analysis of the \textit{AKARI} Monitor field data \citepalias{mat11} for fluctuation of the sky brightness to larger angular scale by using the north ecliptic pole (NEP) deep survey data taken by \textit{AKARI} \citep{hara06}. Compared to the previous study with the Monitor field, the NEP deep field has shorter exposure time but larger areal coverage, enabling us to investigate fluctuations of the sky brightness at larger angular scales up to $\sim$1000$''$. This work is complementary to recent observational studies \citep{coo12b,kash12,zem14} in the sense that we also try to find the existence of the spatial fluctuations at larger angular scale, but at different wavelengths and different regions of the sky. The fluctuation spectra over a wide range of angular scale as well as the spectral shape would be important for the interpretation of the nature of the near-infrared background radiation. 

This paper is organized as follows. In \S2, we present observational data of the NEP deep field, pre-processing of raw image, and other procedures for data reduction. In \S3, we describe power spectrum analysis to measure fluctuations of the sky brightness, and we discuss the possibility of fluctuations by instrumental effects in \S4. Correction of fluctuation spectrum and error estimation are treated in \S5. The contribution of foreground components is described in \S6. In \S7, the result of this study is discussed. We summarize our results in the final section.

\section{OBSERVATIONS AND DATA REDUCTION}

\subsection{Observations}\label{sec:observations}
\textit{AKARI} is an infrared space telescope developed and launched by Japan Aerospace Exploration Agency (JAXA). Details of the mission and its focal plane instruments are described in \citet{mur07} and \citet{Onaka07}. \textit{AKARI} carried out a number of wide area surveys in near-, mid-, and far-infrared wavelength bands. \textit{AKARI} made wide and deep surveys in NEP region \citep{hara06,wad08,kim12}. The survey of the NEP deep field started in 2006 May with all nine wavelength bands of Infra-Red Camera (IRC) that is one of the focal plane instruments (see \citealt{Onaka07} for the detailed description of this instrument) ranging from 2.4 to 24 $\mu$m, and performed about 270 pointed observations over one year period, covering nearly a circular area of 0.4 square degrees (Figure~\ref{fig:imgo}). Among three near-infrared wavelength bands, we used N2 (2.4 $\mu$m) and N3 (3.2 $\mu$m) for this study, as N4 (4.1 $\mu$m) had shorter exposure time than the other ones. Astronomical Observation Template (AOT) of the NEP deep field is IRC05 that performed five sets of exposures (5 images) without dithering for one near-infrared wavelength band during one pointed observation. The exposure time for an image is 65.4 seconds. The field of view (FOV) of an image is about 10$'$ $\times$ 10$'$, and the pixel scale is $1\farcs46$. Further information for observation is available in the IRC data user manual \citep{Lorente08}.

\subsection{Pre-processing}\label{sec:pre-processing}
We used \textit{AKARI} IRC imaging pipeline (version: 070908) for most of the pre-processing of each image \citep{Lorente08}. However, a few procedures, such as dark subtraction, flat field correction, and aspect ratio resampling, were treated separately to improve the quality of the image. In the ordinary pipeline, superdark frame that was made by combining about hundred dark frames taken during Large Magellanic Cloud (LMC) observations is used as a default to subtract dark current. However, ordinary pipeline with superdark could not be used to estimate different enhancement of dark current at each pixel in the array when the telescope was exposed to a large number of cosmic rays. For this reason, \citet{tsu11} studied a new dark subtraction method that is capable of estimating different response of dark current at each pixel even when images were exposed to many cosmic rays.

For the correction of flat field, we made self flat templates using all the NEP deep field images taken during the period when stray light was negligible. About 210 images in the NEP deep field were used to make self flat template for each band. To make self flat template, bright pixels in the image were masked by clipping procedure to remove real objects, cosmic rays, and some prominent artifacts such as multiplexer bleed (MUXbleed hereafter) trails near very bright objects \citep{kim12}. On the other hand, faint pixels that were affected by various kinds of artifacts were treated manually. We examined every image and manually masked faint artifacts such as bad pixels, column pulldown, and center of saturated objects. Eventually, we obtained self flat templates that show improved feature of flat field and describe systematically faint regions in all images. Some stripes, which can induce artificial fluctuations toward the horizontal direction of each image, are seen after applying the correction for the aspect ratio. To prevent such stripe-like artifacts, we adopted the ``nearest'' option instead of ``linear'' option for resampling of pixel values.

\subsection{Correction of Instrumental and Time Varying Components}\label{sec:correction of instrument and time variation components}
After pre-processing, additional correction procedures to remove instrumental effect and zodiacal light contribution are needed. The first additional procedure is related to MUXbleed. Observed images of the NEP deep field contain many bright objects, and some of them induced MUXbleed. MUXbleed manifests as bright stripes along horizontal direction of each image caused by bright objects or strong cosmic rays. MUXbleed stripes do not give any significant problem in this study because they can be easily removed by clipping procedure. However, we found that the pixels in the upper region of MUXbleed trail become fainter compared to those in the lower region. In Figure~\ref{fig:mux_cor}, one example of an image (pointing ID: 2110163) and average pixel value of background for each row is presented to illustrate MUXbleed problem. As shown, there is a large jump in pixel values across the MUXbleed trail. Unfortunately, MUXbleed occurs rather frequently in the NEP deep imaging data. Among the selected images for this study, about 76\% and 78\% for 2.4 and 3.2 $\mu$m, respectively, suffer from the MUXbleed effect. It would be best to exclude all pixels whose values are affected by MUXbleed, but we need to use such pixels in order not to reject too many pixels. For this reason, we tried to correct for MUXbleed rather than abandoning all affected pixels. In Figure~\ref{fig:mux_cor}, left image and graph show a state before correcting for the MUXbleed effect, whereas right parts show a state after the correction. Open circle represents average background value of each row, and dotted line in the graph shows the position of MUXbleed. We assumed that the effect of the MUXbleed reaches the dot-dashed line. Solid line in the left graph is the linear fit of the data points that are thought to have been affected, whereas dashed line is the linear fit of data points that are regarded not to have been affected by MUXbleed. In other words, dashed line is treated as a value of the background of the image when MUXbleed problem does not exist. Correction for the MUXbleed problem was done by using the difference between solid line and dashed line in the left graph. In Section~\ref{sec:contribution of instrumental effects}, we try to confirm the validity of the result of MUXbleed correction by using the area where the effect of MUXbleed is small in the mosaic image.

The second additional procedure is related to the zodiacal light. In order to make an accurate mosaic image, we have to correct for the seasonal variation of the zodiacal light. Because the axis of Earth's orbit is slightly inclined from that of IPD cloud, the zodiacal light towards the NEP shows variation over the observational period. To correct for such effect, we tried to fit the sky brightness by using sinusoidal function with the assumption that the spatial distribution of the zodiacal light is smooth at high ecliptic latitude \citep{kel98,pyo12}. The explicit form for the intensity of the zodiacal light as a function of the Earth's heliocentric ecliptic longitude, $l$, can be written
\begin{equation}
I(l)=a \sin (l-b)+c,
\end{equation}
where parameters $a$, $b$, and $c$ represent amplitude, phase angle, and mean sky brightness, respectively, which can be determined by fitting the observed sky brightness. We used MPFIT\footnote{http://purl.com/net/mpfit} package \citep{mark09} for sinusoidal fitting, and results are presented in Table~\ref{tb:fitting}. In Figure~\ref{fig:zl}, we show the fitting result together with the sky brightness data at two near-infrared bands. We subtracted the fitted function from each image to correct for the time variation of zodiacal light.

The MUXbleed and time variation of zodiacal light may induce artificial spatial fluctuation of sky brightness. Moreover, when we make a mosaic image, sky level of each image is adjusted by using overlapped regions between neighboring images. Therefore, two additional correction procedures that stated above should be applied before making mosaic image.

\subsection{Source Subtraction}\label{sec:source subtraction}
According to \citet{takagi12} and \citet{kato12}, Point Spread Function (PSF) of \textit{AKARI} IRC images is not circular but elongated along a certain direction due to jitters in the satellite pointing, and the direction of elongation varied depending on the position in the mosaic image due to variation of position angle of each image. Therefore, source subtraction should be done before making a mosaic image. Removing procedure of resolved objects in the image consists of two parts: clipping and PSF subtraction. Specific procedures for source removal are almost the same as those of the previous study for the Monitor field \citepalias{mat11}, but we applied additional procedures for saturated bright objects and extended objects to minimize the possible contamination of the relatively faint wing parts of such objects. For saturated objects, sky level of an individual object was estimated using a mean of pixel values within an annulus whose inner and outer radii are proportional to the brightness of the object, and then we additionally applied a circular mask whose radius was extended until residual of the object becomes sky level. For extended objects that were identified by stellarity index in the catalog of optical image \citep{cfht}, we also applied additional circular masks with the same ways for saturated objects.

\subsection{Mosaic Images}\label{sec:mosaic images}
\begin{figure}
\begin{center}
\includegraphics[bb=0 0 270 238]{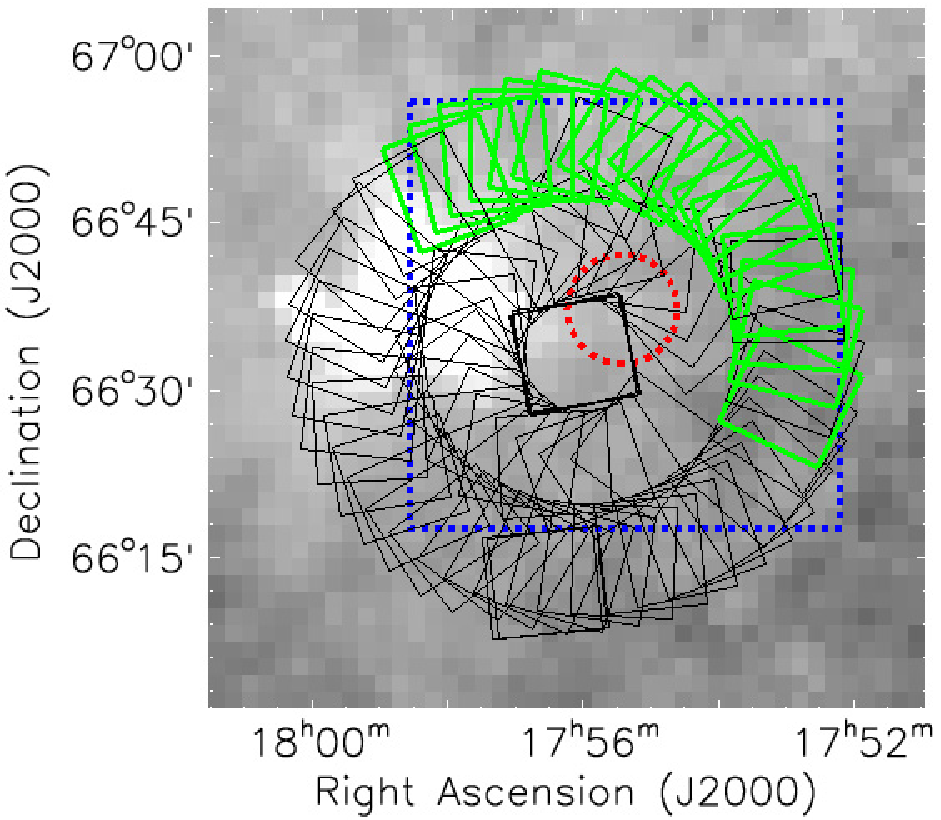}
\caption{Mapping strategy of N2 band in the NEP deep field, which consists of central part, inner annulus, and outer annulus. Solid line box represents FOV of each pointed observation that covers about 10$'$ $\times$ 10$'$. Solid line boxes with thick borders in the outer annulus are the area used in this study, whereas a dotted line circle with a diameter of 10$'$ is the area used in the previous study, i.e., the Monitor field \citepalias{mat11}. A dotted line box shows the position where the simulation for the clustering of unresolved faint galaxies is performed (see Section~\ref{sec:galactic and extragalactic Components}) by using the CFHT WIRCam $K_{s}$ band image, which covers 38$'$ $\times$ 38$'$. Background image is the Improved Reprocessing of the IRAS Survey (IRIS; \citealt{miv05}) 25 $\mu$m map whose FOV corresponds to 64$'$ $\times$ 63$'$, which is retrieved from http://irsa.ipac.caltech.edu/data/IRIS/. Bright (white) region represents a planetary nebula NGC 6543.}\label{fig:imgo}
\end{center}
\end{figure}

\begin{figure}
\begin{center}
\includegraphics[scale=1.0, bb=0 0 424 152]{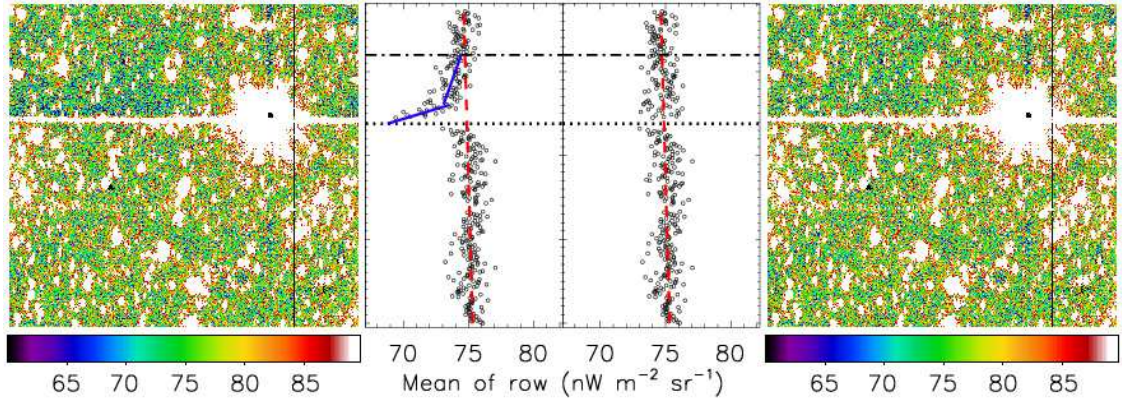}
\caption{Example images (pointing ID: 2110163) of N3 band are introduced to explain the correction for MUXbleed problem. Left image and graph describe a state before MUXbleed correction, whereas right parts describe a state after correction. Open circle represents the average background value of each row. Dotted line represents the position of MUXbleed, and we assume that the effect of MUXbleed reaches to the position where dot-dashed line is presented. Dashed line in the left graph is a fitted line of data points that are regarded not to have been affected by MUXbleed, i.e., it is the base line for the correction. Solid line in the left graph is a fitted line of data points that are thought to have been affected by MUXbleed. Correction for MUXbleed problem was performed by using the difference between solid line and dashed line in the left graph.}\label{fig:mux_cor}
\end{center}
\end{figure}

\begin{figure}
\begin{center}
\includegraphics[scale=0.8]{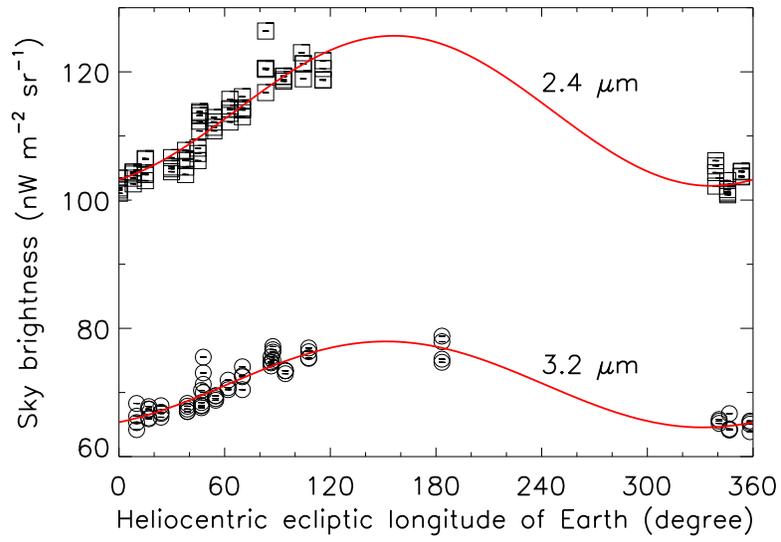}
\caption{Sky brightness of each image for two bands. Open squares and open circles correspond to 2.4 and 3.2 $\mu$m, respectively. Solid lines are sinusoidal fitting of the sky brightness as a function of heliocentric ecliptic longitude of Earth. Error bar represents 1$\sigma$ error.}\label{fig:zl}
\end{center}
\end{figure}

\textit{AKARI} NEP deep survey observations started in 2006 May, but data taken in the northern sky during May, June, and July 2006 were affected by scattered Earth shine \citep{wad07}. For this reason, we excluded the data taken between April and August to avoid the Earth shine problem. We selected 86 pointed observations for two bands taken from 2006 September to 2007 March. We further rejected additional images to obtain a reliable result of this study. First, images containing the planetary nebula NGC 6543 (the bright region in Figure~\ref{fig:imgo}) were rejected because it is not easy to subtract the contribution of a bright and irregular object. Second, we did not use images in the inner annulus but used only those in the outer annulus because of the difficulties in making a reliable mosaic image that includes both inner and outer annuli. To obtain a stable and reliable mosaic image, adjacent images should share sufficient FOV. As shown in Figure~\ref{fig:imgo}, however, the overlapping region between inner and outer annuli is too small for that. Third, bad images that show poor stability during observation or various artifacts including cosmic rays were also rejected. Eventually we used 75 and 68 images for 2.4 and 3.2 $\mu$m, respectively, out of 34 pointed observations. These images after source subtraction with pre-processing and two additional corrections were combined to make mosaic images for two bands by using the software Montage\footnote{http://montage.ipac.caltech.edu/ for more information about this software.}. Montage adjusts sky level of each image by using overlapping regions among adjacent images during mosaicking. Because sky level of an image is usually not exactly consistent with that of the neighboring image in the overlapping region, sky adjustment procedure while making mosaic images further reduces artificial fluctuations.

\begin{figure}
\begin{center}
\includegraphics[scale=1.0, bb=0 0 289 279]{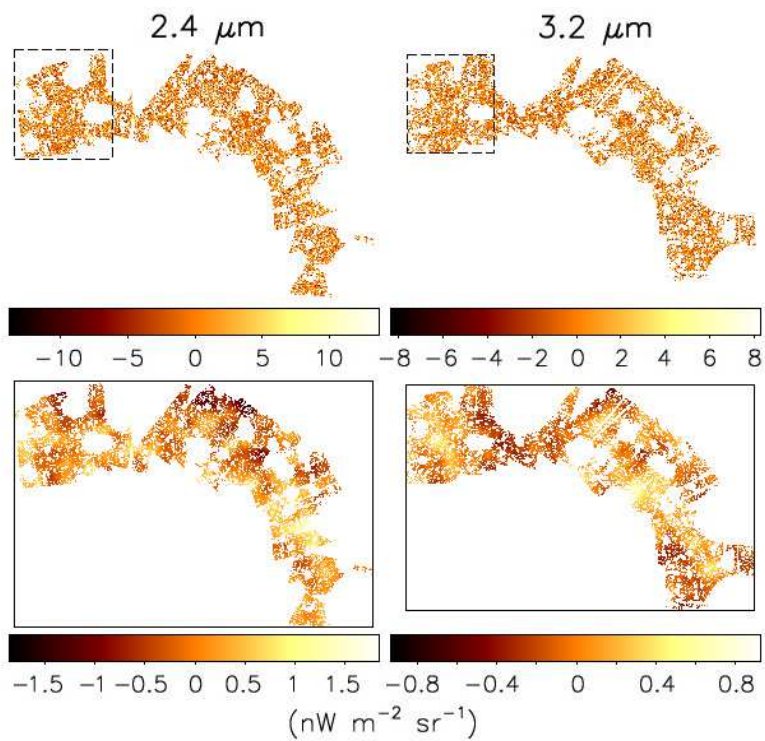}
\caption{Mosaic images of the NEP deep field for two bands in the equatorial coordinates. Top: final states of the mosaic images. White color represents masked regions. The small area denoted by dashed line box is the region where the effect of MUXbleed is small compared to other regions in the mosaic image regardless of the correction for MUXbleed problem. Bottom: smoothed mosaic images whose pixel values are replaced by non-weighted averaging of neighbor pixels within a circular shape filter with a diameter of 100$''$ to show large scale structures. The size of the mosaic image (solid line box) is 36$'$ $\times$ 25$'$ and 35$'$ $\times$ 23$'$ for 2.4 and 3.2 $\mu$m, respectively. Note that structures in the mosaic images include artificial as well as real ones in the sky. Mean pixel value of each mosaic image is adjusted to zero.}\label{fig:imgf}
\end{center}
\end{figure}

In order to guarantee a stable mosaicking procedure, sufficient overlapping regions among adjacent images are necessary. For this reason, although source subtraction procedure was performed before mosaicking, further masking for saturated bright objects and extended objects was conducted in the mosaic image to leave sufficient overlapping regions for mosaicking. Moreover, instrumental and observational artificial components, such as bad pixels, column pulldown, center of the saturated objects, and MUXbleed, were removed in advance before mosaicking because they hinder sky matching process because of their different pixel values and shapes in the adjacent images. After mosaicking, pixels that remain at least in two pointed observations are selected for the study (the top panel of Figure~\ref{fig:imgf}). Because pixels that are survived only in a single pointed observation do not have a chance to suppress the contributions of artificial components, they are more likely to be affected by artifacts such as leakage of stray light (see Section~\ref{sec:contribution of instrumental effects} for details). In addition, it is hard to recognize whether the boundary region of an image is affected by objects just outside of the image. In order to avoid such contamination, we selected pixels that are also survived at least in an additional single adjacent pointed observation. 

The small area denoted by a dashed line box in Figure~\ref{fig:imgf} represents the region where the effect for MUXbleed is small compared to the other regions because this area does not have any bright object. We use this small area to confirm the validity of the correction for MUXbleed problem. We can measure fluctuations of the sky brightness by using mosaic images. Basic parameters of the mosaic image of the NEP deep field after the correction procedures are listed in Table~\ref{tb:data_deep}. In the bottom panel of Figure~\ref{fig:imgf}, smoothed mosaic images made by averaging of neighbor pixels within a circular shape filter with a diameter of 100$''$ are presented to show spatial structure more clearly. Although, spatial structure is seen in the smoothed mosaic images, its pattern appears to be somewhat different for each band. This could imply a possibility of the existence of artificial structure. More careful assessment of such phenomenon is discussed later in Section~\ref{sec:contribution of instrumental effects}.

\section{POWER SPECTRUM ANALYSIS}

\begin{figure}
\begin{center}
\includegraphics[scale=0.8]{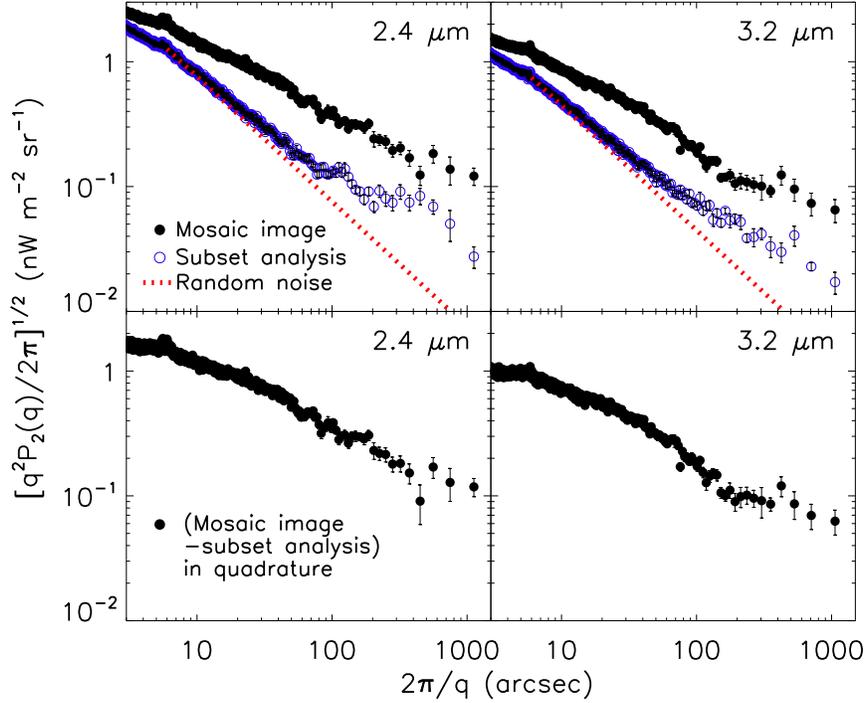}
\caption{Top: fluctuation spectra, [$q^{2}P_{2}$($q$)/2$\pi$]$^{1/2}$, of the mosaic image (filled circles) and subset analysis (open circles), respectively. Dotted line represents the expected slope from random noise. Bottom: fluctuation spectra of the mosaic images after quadratically subtracting those of subset analysis. $P_{2}$($q$) and $q$ represent power spectrum and angular wavenumber, respectively. Error bars show 1$\sigma$ error.}\label{fig:final}
\end{center}
\end{figure}

We used power spectrum analysis to measure spatial fluctuation of the NEP deep field following the previous works (refer to \citetalias{mat11} for the more detailed description of power spectrum analysis). The top panel of Figure~\ref{fig:final} represents the fluctuation spectra, [$q^{2}P_{2}$($q$)/2$\pi$]$^{1/2}$, for the mosaic images (filled circles) and subset analysis (open circles). Filled circles contain contributions of instrumental noise and unresolved faint objects as well as background component. We estimated the contribution of each component separately to extract background component. Subset analysis is intended to estimate fluctuation spectrum of instrumental noise in the mosaic image and to confirm whether observed structure in the mosaic image is of celestial or cosmic origin. To perform subset analysis, the whole images were divided into two subsets composed of alternating pointed observations in sequence from the left side of the mosaic image, and a difference of two subsets was taken. The difference of two subsets will erase persistent features in the sky, and leave instrumental noise that consists of random noise and artificial feature in the mosaic image. As shown in the top panel of Figure~\ref{fig:final}, the subset analysis reveals significantly smaller fluctuation than that of the mosaic image. However, the result of subset analysis shows larger fluctuation than that of random noise at large angular scale. As discussed later in Section~\ref{sec:contribution of instrumental effects}, this implies that there could remain artificial structure in the mosaic image. A difference of fluctuation spectra between the subset analysis and random noise represents the amount of the contamination by the possible artificial structure. At angular scales larger than 100$''$, a mean power spectrum, $P_{2}$($q$), due to artificial structure reaches up to about 10\% and 12\% of that of the mosaic image for 2.4 and 3.2 $\mu$m, respectively. Fluctuation spectra of the mosaic images after quadratically subtracting the results of the subset analysis are presented in the bottom panel of Figure~\ref{fig:final}.

\section{CONTRIBUTION OF INSTRUMENTAL EFFECTS}\label{sec:contribution of instrumental effects}

\begin{figure}
\begin{center}
\includegraphics[scale=0.8]{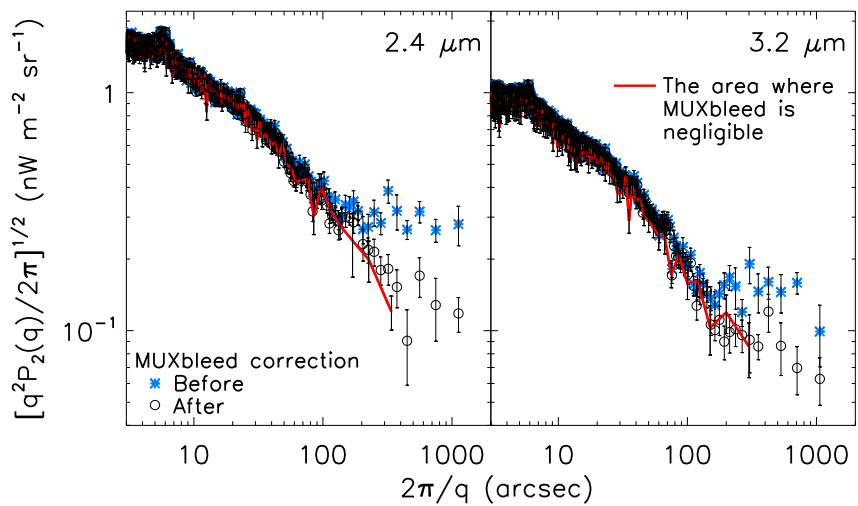}
\caption{Fluctuation spectra of the mosaic images before (asterisks) and after (open circles) MUXbleed correction for 2.4 and 3.2 $\mu$m, respectively. Solid line is fluctuation spectrum at the small area where the effect of MUXbleed is negligible compared to other regions in the mosaic image. We can see that the fluctuation spectrum after the MUXbleed correction is very similar to that of the area where MUXbleed is negligible up to $\sim$300$''$. Error bar represents 1$\sigma$ error.}\label{fig:compare}
\end{center}
\end{figure}

One of the possible origins of the spatial fluctuations of the infrared background is the instrumental effect. MUXbleed effect could induce artificial fluctuation. In order to check any possible artificial fluctuations due to insufficient correction for MUXbleed, we used the small area (denoted by dashed line box in Figure~\ref{fig:imgf}) in the mosaic image as a representative region where the effect of MUXbleed problem is negligible even before MUXbleed correction. In Figure~\ref{fig:compare}, two new fluctuation spectra (solid lines and asterisks) are shown along with the spectra derived in the previous section (open circles). Solid lines represent fluctuation spectra of the small areas where the effect of MUXbleed is negligible compared to other regions in the mosaic image. In addition, we made another kind of mosaic image for each band composed of images before the correction of MUXbleed. Their fluctuation spectra are presented as asterisks. Because the effect of MUXbleed is not confined in a single image but propagates to all regions during the mosaicking procedure, asterisks show larger fluctuations than open circles at angular scales larger than 100$''$. Moreover, open circles have similar fluctuation power to that of solid lines at angular scales up to a few hundreds arcseconds, implying that significant amount of MUXbleed problem has been removed by the correction procedure shown in Figure~\ref{fig:mux_cor}. We can see that the whole mosaic image after MUXbleed correction shows similar fluctuation power to the small area where the effect of MUXbleed is negligible. We conclude that the correction for MUXbleed problem was effective.

\begin{figure}
\begin{center}
\includegraphics[scale=0.8]{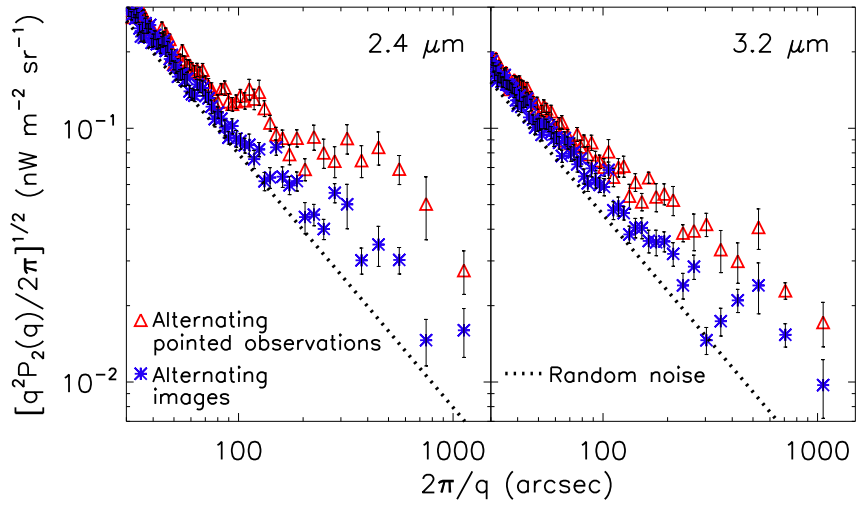}
\caption{Fluctuation spectra of the subset analysis with alternating pointed observations (open triangles) and alternating images (asterisks) for 2.4 and 3.2 $\mu$m, respectively. Error bar is 1$\sigma$ error. Dotted line represents the expected slope from random noise.}\label{fig:subset}
\end{center}
\end{figure}

Meanwhile, as shown in the top panel of Figure~\ref{fig:final}, the result of subset analysis gives larger fluctuation than that of random noise at large angular scales. This means that subsequent pointed observations toward the same field show different structures with each other, which may have been originated from temporary instrumental effects. Although contributions of artificial components are estimated and subtracted from the final result by subset analysis, we attempted to identify the cause of large fluctuation in subset analysis to ensure the result of this study. We made two subsets that contain alternating images (i.e., alternating exposures in a pointed observation), rather than alternating pointed observations, in sequence from the left side of the mosaic image. As presented with asterisks in Figure~\ref{fig:subset}, the result of new subset analysis shows smaller fluctuation than the previous one and becomes closer to the fluctuation of random noise shown as a dotted line at large angles. This implies that there exist artificial structures in each pointed observation, which persist at least during a period of one pointed observation. In the new subset analysis, an image in one subset shares the same FOV with its counterpart in the other subset due to no dithering between them. Therefore although there are artificial structures in each image, they are canceled out each other by taking a difference of two subsets if artificial structures are unchanged during a period of one pointed observation. As a result, new subset analysis shows nearly random-noise-like fluctuation spectrum.

There are several candidates that can induce artificial structures in each pointed observation. First, errors occurred during pre-processing of raw imaging data can be a source of the artificial structure. However, because we used the same pre-processing method with that of the previous study \citepalias{mat11} and the result of subset analysis in the previous study shows random-noise-like behavior, the problems in pre-processing are not likely to be the cause. Another possible origin is the error that arises during source subtraction procedures by using PSF. Because PSF of \textit{AKARI} IRC image is not symmetric but elongated and the direction of PSF elongation in the mosaic image varies depending on the position angle of the pointed observation, there is a possibility of distinct residuals in each pointed observation after PSF subtraction. If so, the effect of PSF subtraction may produce artificial fluctuation in subset analysis at small angular scales of a few tens arcseconds as well as at large angular scales. As shown in Figure~\ref{fig:subset}, the large fluctuation in subset analysis mainly appears at angular scales larger than 100$''$, whereas the result of subset analysis follows the behavior of random noise at small angular scales. Therefore, the error related with PSF subtraction procedure is not likely to be the main reason for the large fluctuation at large angular scale in subset analysis. 

\begin{figure}
\begin{center}
\includegraphics[scale=0.9, bb=0 0 383 480]{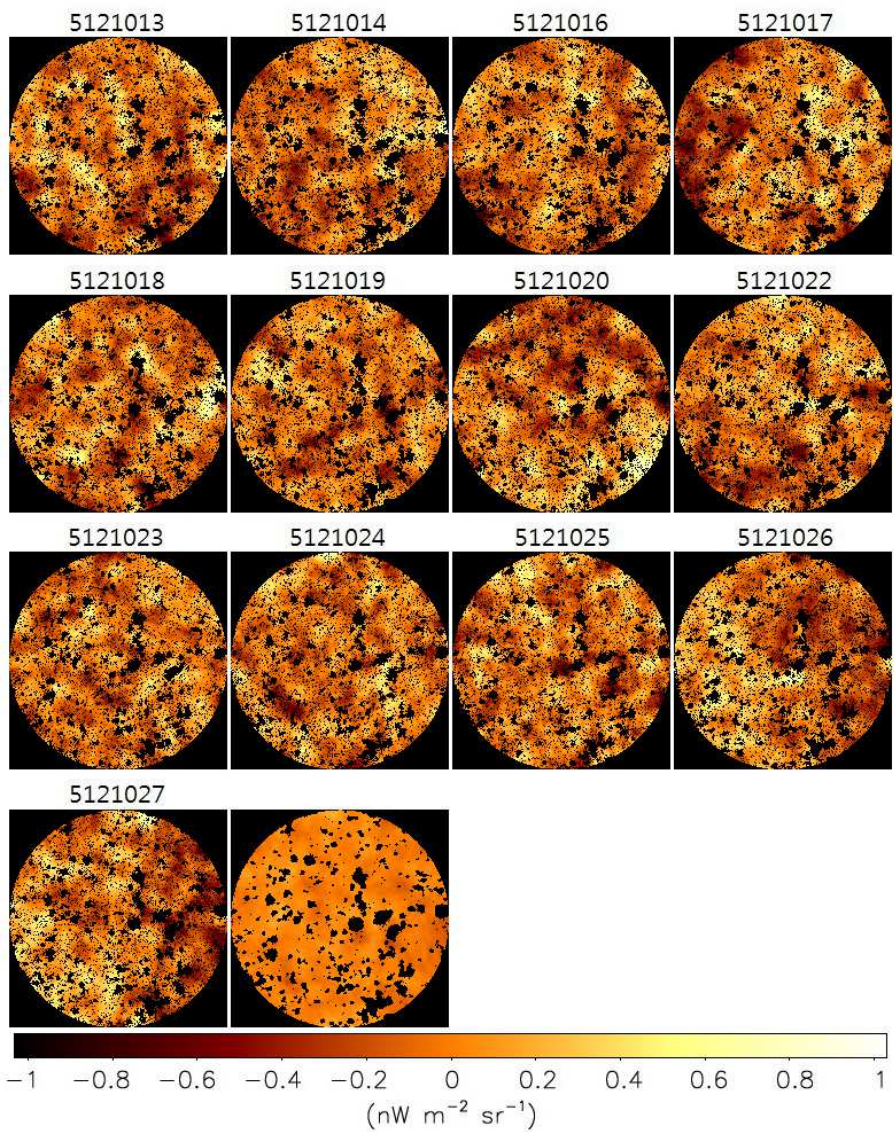}
\caption{Here we show 13 images (N2 band) with pointing IDs (numbers above images) that are produced by subtracting the stacked image of each pointed observation of the Monitor field separately from the entire stacked image of the Monitor field. The last image is the stacked image of 13 images with pointing IDs. Images are smoothed by non-weighted averaging of neighbor pixels within a circular shape filter with a diameter of 50$''$ to emphasize structures. Black color represents masked regions. The diameter of each image is about 550$''$. Images are aligned with the equatorial coordinates. Mean pixel value of each image is adjusted to zero. Scale range corresponds to $\pm$5$\sigma$ of 13 images with pointing IDs.}\label{fig:monitor}
\end{center}
\end{figure}

The remaining probable cause is temporary instrumental effects such as stray light that remain as distinct structures in each pointed observation. We paid particular attention to the fact that stray light can change during a period of subsequent pointed observations. To test the stray light as a possible origin of the artificial fluctuation, we made more detailed analysis of the imaging data of the Monitor field. We found 13 pointed observations of the Monitor field that were carried out during the same period, and 13 images were made by stacking 3 images of N2 band in each pointed observation separately. After that, 13 images were subtracted from the entire stacked image of N2 band in the Monitor field separately. We present result in Figure~\ref{fig:monitor}, which shows distinct structure in each pointed observation. They are artificial structures because real structure is removed by subtraction of the entire stacked image of the Monitor field. The last image in Figure~\ref{fig:monitor} was made by stacking 13 images, which shows negligible structure and indicates that artificial structures in each pointed observation have random nature compared to those in other pointed observations. The test with the Monitor field data has some advantages in identifying possible candidates that could lead to the large fluctuation in subset analysis. For example, because they do not contain MUXbleed problem and subset analysis of the Monitor field shows similar behavior with random noise \citepalias{mat11}, we can exclude MUXbleed problem and errors occurred during pre-processing from the possible candidates. In addition, because only clipping procedure was used to remove objects in this test, we can also exclude errors occurred during PSF subtraction procedure as a possible candidates.

\begin{figure}
\begin{center}
\includegraphics[scale=0.8]{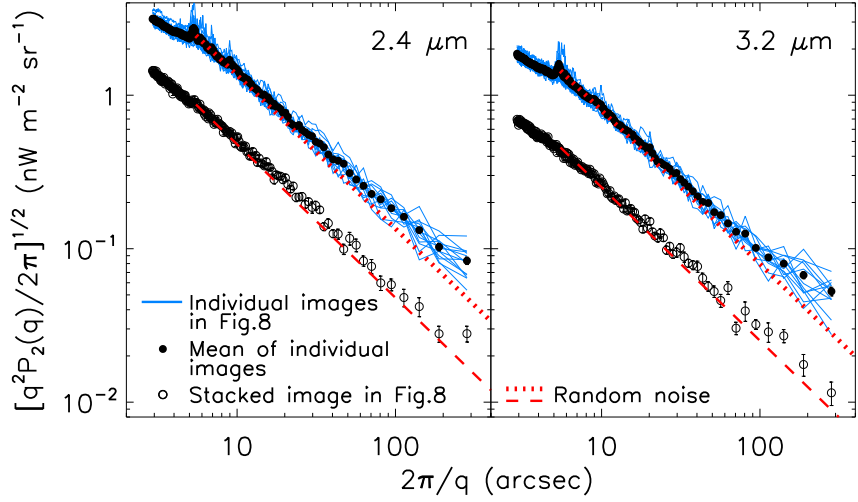}
\caption{Fluctuation spectra of the individual images (13 images with pointing IDs in Figure~\ref{fig:monitor}; solid lines) and the stacked image (the last image in Figure~\ref{fig:monitor}; open circles) for 2.4 and 3.2 $\mu$m, respectively. Filled circles represent mean of the solid lines. Dotted line and dashed line represent the expected slope from random noise for each case. All fluctuation spectra are estimated by using images before smoothing. Error bars show 1$\sigma$ error. Error bars of the solid lines are not plotted to avoid confusion, while those of the filled circles are too small to be seen.}\label{fig:monitor_ps}
\end{center}
\end{figure}

We made same test for N3 band to examine the behavior of artificial structures in the Monitor field. Note that AOT of the Monitor field is IRC03 that is different with that of the NEP deep field. IRC03 performed three sets of images for three near-infrared wavelength bands of the IRC alternately during one pointed observation, and filter wheel change or dithering was applied between exposures. We also found artificial structures in N3 band and obtained linear Pearson correlation coefficients of 0.112(mean)$\pm$0.158(standard deviation) with those of N2 band in each pointed observation. Evidently there is no statistically significant correlation between them even in each pointed observation. That is to say, unlike artificial structures in the NEP deep field, those in the Monitor field vary frequently even during a period of one pointed observation. A probable candidate that can explain both the existence of artificial structures in two fields and the difference of their behaviors is a contribution of stray light that depends on the position of filter wheel. Although we used imaging data that were observed during a period of small stray light, we could not exclude some leakage of stray light that has contaminated the images. In the case of the NEP deep field, because there is no operation of filter wheel during a pointed observation, the pattern of the stray light remained nearly the same for different images taken during one pointed observation. On the other hand, in the case of the Monitor field, because stray light could have been scattered differently due to operation of filter wheel during a pointed observation, we can expect different shape of artificial structure in each image. Therefore, distinct artificial pattern may remain in each pointed observation of the NEP deep field more clearly than that of the Monitor field. Moreover, such artifacts are rarely suppressed during mosaicking due to small number of stacked images per field. To evaluate the contribution of artificial structures quantitatively, we estimated fluctuation spectra of the images in Figure~\ref{fig:monitor}. The results for 13 pointed observations are presented as solid lines in Figure~\ref{fig:monitor_ps}, and a mean of them is presented as filled circles, which show larger fluctuation than dotted line at large angular scales. Fluctuation spectrum of the last image denoted by open circles shows small deviation from random noise and smaller overall fluctuation power than those of other images.

The amount of contribution of artificial structure in the mosaic image of the NEP deep field can be estimated by quadratically subtracting the fluctuation spectrum of random noise from that of subset analysis in the top panel of Figure~\ref{fig:final}, which becomes 0.098 and 0.053 nW m$^{-2}$ sr$^{-1}$ for 2.4 and 3.2 $\mu$m, respectively, as a mean at angles between 100$''$ and 350$''$. We can expect such quantitative contribution of artificial structure of the NEP deep field by converting that of the Monitor field presented in Figure~\ref{fig:monitor_ps}. A difference between filled circle and dotted line represents fluctuation due to artificial structure in each pointed observation of the Monitor field. For the case of N2 band, a mean difference at angular scales larger than 100$''$ is about 0.092 nW m$^{-2}$ sr$^{-1}$. In a pointed observation of the Monitor field, an amplitude of artificial structure becomes $\sqrt{3}$ times weaker as 3 images with different shape of artificial structure are stacked with dithering. However, such a reduction would not occur for the NEP deep field because images with similar shape of artificial structure are stacked without dithering. As a result, in the NEP deep field, we can expect that the contribution of artificial structure in each pointed observation is $\sqrt{3}$ times larger than that of the Monitor field. Finally, because the mosaic image for N2 band in the NEP deep field is covered by 3.4 pointed observations per field on average, the contribution of artificial structure becomes $\sqrt{3.4}$ times weaker in the mosaic image, resulting in 0.086 nW m$^{-2}$ sr$^{-1}$. Similarly we estimated the fluctuation due to artificial structure to be 0.049 nW m$^{-2}$ sr$^{-1}$ for N3 band in the NEP deep field. Such result corresponds to about 90\% of the residual fluctuation of subset analysis over random noise shown in the upper panel of Figure~\ref{fig:final}. Therefore we conclude that the contribution of temporary instrumental effects, most likely stray light, may be the most prominent cause for the large fluctuation in subset analysis. Although this test is valid only within angular scales of a few hundreds arcseconds, it can give a hint for the large fluctuation of subset analysis at angular scales larger than a few hundreds arcseconds.

Although we examined artificial components in the mosaic image and their possible origin, it is difficult to correct them more accurately due to their random nature in each pointed observation. To reduce further the contribution of artifacts with random nature, we need larger number of stacked images per field.

\section{CORRECTION AND ERROR ESTIMATION}\label{sec:correction and error estimation}

\subsection{Corrections of the Fluctuation Spectrum}\label{sec:corrections of the fluctuation spectrum}
Fluctuation spectra in the bottom panel of Figure~\ref{fig:final} were further corrected to remove instrumental and systematic effects. Corrections of the fluctuation spectra were conducted as follows. The power spectrum after the corrections can be expressed as
\begin{equation}
P_{2}(q) = \frac{M^{-1}(q)\tilde{P}_{2}(q)}{T^{2}(q)B(q)},
\end{equation}
where $M(q)$ is a mode coupling matrix, $T(q)$ is a map-making transfer function, $B(q)$ is a beam transfer function, and $\tilde{P}_2(q)$ represents power spectrum before the corrections.

\begin{figure}
\begin{center}
\includegraphics[scale=0.8]{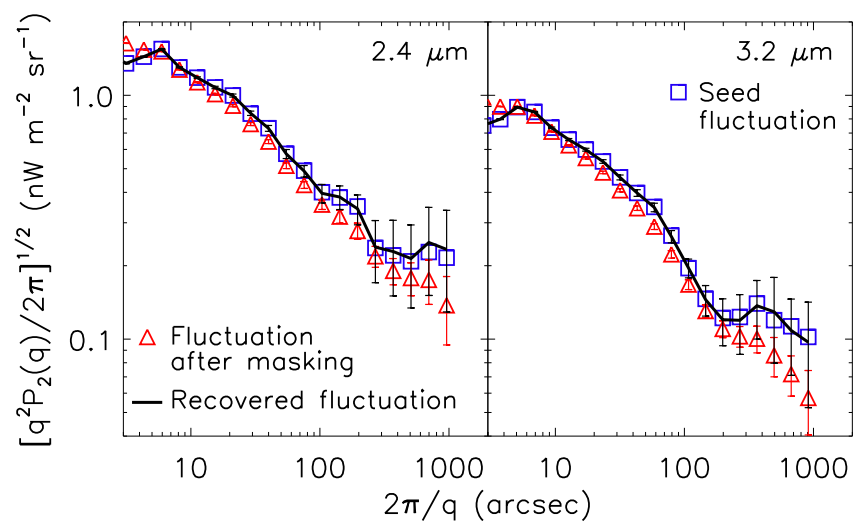}
\caption{Test of the mode coupling matrix for 2.4 and 3.2 $\mu$m, respectively. Squares are seed fluctuations for simulations, and triangles are fluctuation spectra of the simulations after adopting mask of the mosaic image. Solid lines are recovered fluctuations by the inverse of the mode coupling matrix, $M^{-1}(q)$. Symbols and error bars represent mean and standard deviation of 500 simulations.}\label{fig:mc_ps}
\end{center}
\end{figure}

Observed fluctuation is affected by the presence of mask in the mosaic image. In order to correct the observed fluctuation, we examined an effect of the mask on each data point separately by simulation \citep{coo12b,zem14}. We made a simulated image that has a unit value of fluctuation power at a certain angular scale but no fluctuations at other angular scales. By adopting mask of the mosaic image on the simulated image, we can estimate how the mask converts the unit fluctuation power at a certain angular scale into those of other angular scales. Simulations were iterated 100 times for each data point by adjusting phase of a complex number in Fourier space, and a result was expressed as a matrix, $M(q)$, whose each row corresponds to the result of simulation for each data point. The mode coupling matrix can be used to estimate the effect of mask on the fluctuation, whereas its inverse, $M^{-1}(q)$, can be used to recover the intrinsic fluctuation. To check usefulness of the mode coupling matrix, we introduced a seed fluctuation that has similar fluctuation power to the observation and produced 500 simulated images with identical fluctuation spectrum to the seed fluctuation. Each simulated image corresponds to a random realization of the seed fluctuation by adjusting phase in Fourier space. Fluctuation spectra of the simulated images with mask of the mosaic image were recovered by the inverse of the mode coupling matrix. As shown in Figure~\ref{fig:mc_ps}, mean of recovered fluctuations (solid line) is well matched with the seed fluctuation (open square). By using simulated mask pattern, we confirmed that an error of the recovered fluctuation is related with the portion of the mask. Because FOV of the NEP deep field is not rectangular, outside of the FOV is also treated as a mask in the simulation. Therefore such large portion of the mask induces large error of the recovered fluctuation. Figure~\ref{fig:mc_ps} shows that the mode coupling matrix is efficient in estimating the mean effect of the mask for multiple realizations. Because there is no information about an intrinsic fluctuation spectrum of the NEP deep field, we do not know how the intrinsic fluctuation spectrum of the NEP deep field is to be converted by mask of the mosaic image. Therefore, we adopted the mean effect of the mask for the correction. Two components are considered for the error of the mode coupling matrix. One is the systematic error that can be measured by difference between seed fluctuation and recovered fluctuation in Fig. 10. The other is the statistical error as measured by the standard deviation of the recovered fluctuations for many different realizations. Two components are added quadratically, and it becomes the most dominant component of the error budget at angular scales larger than 300$''$. However, at that angular scale, mean errors of the mode coupling matrix for 2.4 and 3.2 $\mu$m, 0.086 and 0.054 nW m$^{-2}$ sr$^{-1}$, are smaller than observed fluctuations for 2.4 and 3.2 $\mu$m, 0.177 and 0.129 nW m$^{-2}$ sr$^{-1}$.

\begin{figure}
\begin{center}
\includegraphics[scale=0.8]{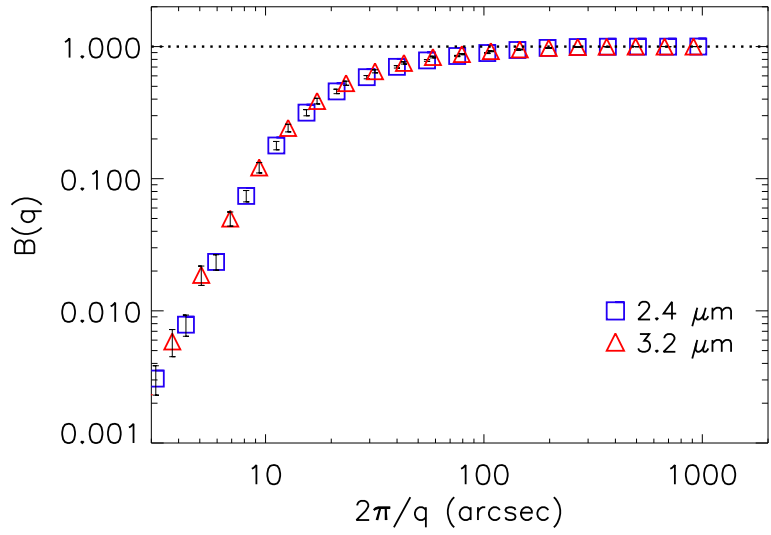}
\caption{The beam transfer function of the NEP deep field. Symbols represent power spectra of PSFs that are normalized to be 1 at large angular scale. Squares and triangles are results for 2.4 and 3.2 $\mu$m, respectively. Error bars represent a standard deviation of PSFs in pointed observations.}\label{fig:bc}
\end{center}
\end{figure}

Fluctuation power at small angular scale can be reduced due to the effect of PSF. Although we used analytic form for the PSF to subtract wing parts of detected objects, it is difficult to define asymmetric central parts of objects by using the analytic function. Therefore, to correct for the effect of PSF on the fluctuation, we derived PSF of the NEP deep field by stacking point objects. We stacked 127 and 85 point objects for 2.4 and 3.2 $\mu$m, respectively, to obtain PSFs and their power spectra of these bands. The beam transfer function in Figure~\ref{fig:bc} represents power spectrum of PSF that is normalized to be 1 at large angular scale, and error bar represents a variation of PSF in pointed observations. Because PSF of one pointed observation is a little different from those of other pointed observations due to variation of jitters in the satellite pointing, we modeled PSF and estimated its power spectrum separately for each pointed observation. Error bar in Figure~\ref{fig:bc} represents their standard deviation. To correct for the contribution of PSF, observed power spectrum was divided by the beam transfer function. The error was included in the final error budget after applying error propagation formula \citep[page 43]{bev92}. At angular scales larger than 100$''$, the error related to the beam transfer function becomes smaller than 0.002 and 0.001 nW m$^{-2}$ sr$^{-1}$ for 2.4 and 3.2 $\mu$m, respectively, which are smaller than observed fluctuations by two orders of magnitude and negligible compared to other error components.

\begin{figure}
\begin{center}
\includegraphics[scale=0.8]{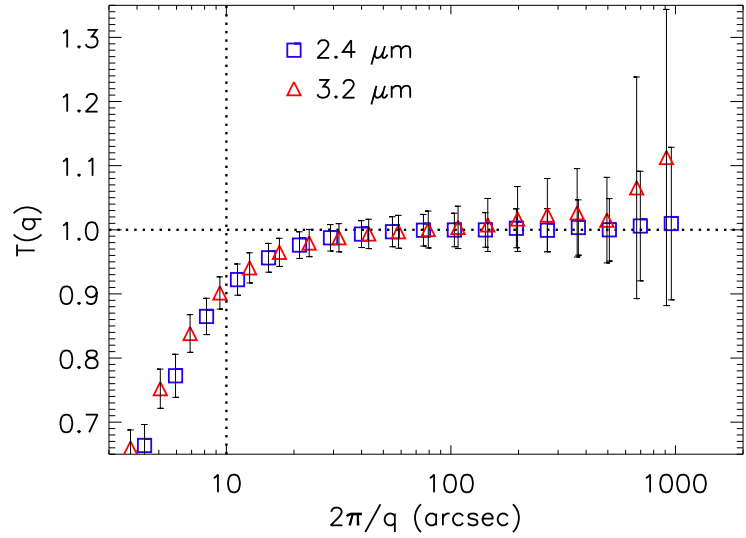}
\caption{The map-making transfer function by the mosaicking procedure. It shows a ratio of remade fluctuation spectra to original ones. Squares and triangles are results for 2.4 and 3.2 $\mu$m, respectively. Symbols and error bars represent mean and standard deviation of 1000 simulations. We used observed fluctuation spectra at angular scales larger than 10$''$ for further analysis because suppression of the fluctuation becomes significant at angles smaller than 10$''$.}\label{fig:mtf}
\end{center}
\end{figure}

In order to test a reliability of the mosaicking procedure, one usually uses a map-making transfer function, $T(q)$ \citep{coo12b}. First, we generated an image of the sky that is large enough to cover the analyzed portion of the NEP deep field in this study (36$'$ $\times$ 25$'$ and 35$'$ $\times$ 23$'$ for 2.4 and 3.2 $\mu$m, respectively) with fluctuation power similar to the observed one in order to mimic the real sky. Individual frames identical to the size of \textit{AKARI} IRC frame have been extracted from the simulated image. We then assigned the same noise and sky levels, and applied the same masking patterns with observed frames to all extracted frames. An intensity of the random noise in the simulated image was determined by the result of subset analysis at small angular scale, and sky level of each simulated image was determined using the value in Figure~\ref{fig:zl}. Finally, they were combined through the same mosaicking procedure using Montage as we did for the observed images. Simulations were repeated for 1000 times, and original fluctuation spectra were compared with those of remade mosaic images. The map-making transfer function in Figure~\ref{fig:mtf} represents mean of the ratio of remade fluctuation spectra to original ones, and error bar represents a standard deviation of 1000 simulations. First of all, at large angular scale, we were able to reproduce the original fluctuation for 2.4 $\mu$m, while we found fluctuation larger by about 10\% at the largest angular scale for 3.2 $\mu$m. One possible cause of such a difference is difference in overlapping regions between 2.4 and 3.2 $\mu$m: there are more overlapping region in 2.4 $\mu$m than in 3.2 $\mu$m. However, this is mainly due to the order of data reduction procedure in this study. Because we masked some pixels before making the mosaic image to remove objects, overlapping region between adjacent images becomes small. It may be a main reason that for additional artificial fluctuations at large angular scale for 3.2 $\mu$m. To confirm this, we obtained the map-making transfer function for 3.2 $\mu$m without masking before making the mosaic image. At large angular scale, we found that the additional fluctuation is reduced to about 2\% of the original fluctuation for 3.2 $\mu$m. Also the standard deviation of 1000 simulations decreased by about one third compared to the result in Figure~\ref{fig:mtf}. On the other hand, significant suppression in the fluctuation spectrum is apparent at small angular scale due to smoothing effect by resampling of pixel values during mosaicking procedure. We found that such suppression is mainly induced by instrumental random noise rather than celestial objects at angular scales smaller than 10$''$. Therefore, we did not use the fluctuation spectrum at angular scales smaller than 10$''$ for further analysis since the measured fluctuations at such small scales are dominated by the shot noise and lie beyond our interests. To correct for the effect of the map-making transfer function, observed fluctuation spectrum was divided by the map-making transfer function. The error was included in the final error budget after applying the error propagation formula. At angular scales larger than 100$''$, mean of the error related to the map-making transfer function is 0.009 and 0.006 nW m$^{-2}$ sr$^{-1}$ for 2.4 and 3.2 $\mu$m, respectively, which are smaller than observed fluctuations by an order of magnitude.

\subsection{Systematic Errors}\label{sec:systematic errors}
\begin{figure}
\begin{center}
\includegraphics[scale=0.8]{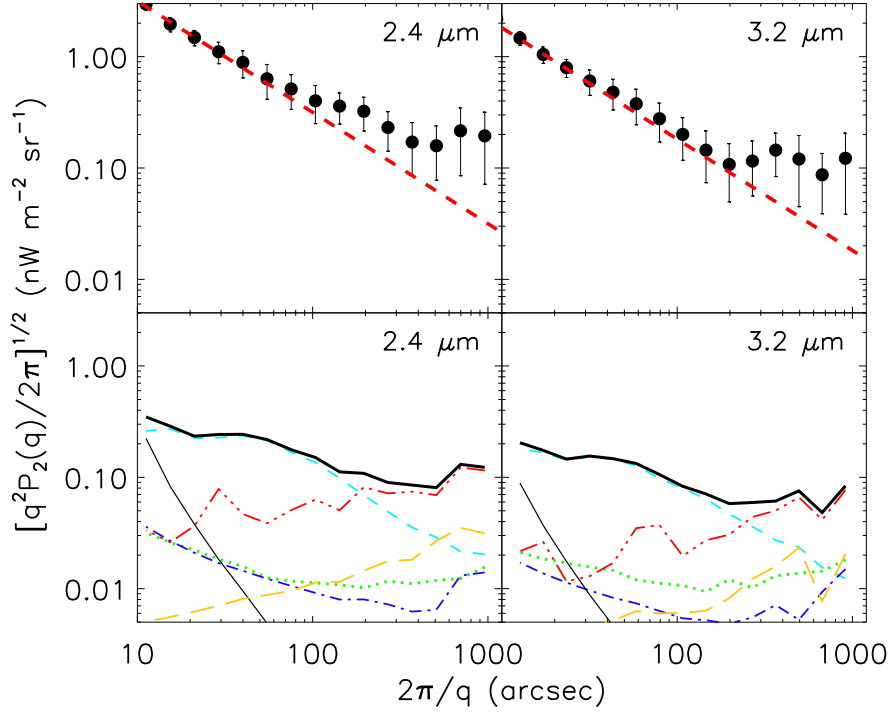}
\caption{Top: the final fluctuation spectra (filled circles) for 2.4 and 3.2 $\mu$m, respectively. Dashed line represents the contribution of shot noise that is estimated by simulations (see text for details). Error bar of the filled circle represents quadratic sum of all error components that are presented in the bottom panel. Bottom: all components of the error are plotted separately by various lines: 1$\sigma$ error of Fourier elements in each bin (long dashed), flat field error (dotted), residuals of detected objects (dashed), and errors related with the mode coupling matrix (triple dot dashed), the beam transfer function (thin solid), and the map-making transfer function (dot dashed). Thick solid line is quadratic sum of them.}\label{fig:real_final}
\end{center}
\end{figure}

In the previous section, errors related with the mode coupling matrix, the beam transfer function, and the map-making transfer function were described. In addition, flat field error and an error due to residual of detected objects were investigated as possible systematic errors because their contributions to the observed fluctuation spectrum may be significant. Even a small amount of flat field error in an image can be accumulated in the mosaic image during mosaicking procedure. Therefore, artificial fluctuation due to flat field error can affect the final fluctuation spectrum. In order to obtain flat field error, we stacked 190 and 220 images for 2.4 and 3.2 $\mu$m, respectively, with ($x$, $y$) coordinate after masking detected objects. After that, the obtained flat pattern is assigned to individual frames for each band and mask pattern and sky level were also assigned to them. Finally, they were combined through the same mosaicking procedure using Montage as we did for the observed images, and fluctuation spectra were estimated. The result of this simulation can be treated as an upper limit due to flat field error because simulated image also contains celestial fluctuation and fluctuations due to other artifacts. As shown in Figure~\ref{fig:real_final}, fluctuation spectra of the simulated images are smaller than observed fluctuations at all angular scales, indicating that flat field error is not significant and not severely accumulated in the mosaic image during mosaicking procedure.

Residuals after masking and subtraction of detected objects can also give some contributions to the final result. If there remain residuals of detected objects, their contributions are most significant just outside of the masked region of the objects. For this reason, we examined residual within 5 pixel layers from the masked region by comparing sky level of each detected object. After that, to make a simulated image with a size of the mosaic image, mean of residual of each detected object is assigned to that region, while zero value is assigned to other regions. Instead of real residual, mean of residual is assigned to suppress the contribution of noise and undetected faint object in that region. As shown in Figure~\ref{fig:real_final}, the contribution from the residual of detected objects is also smaller than observation especially at large angular scale, implying that the observed excess fluctuation at large angular scale is not affected by the residual of the detected objects.

\subsection{Corrected Fluctuation Spectrum}\label{sec:corrected fluctuation spectrum}
Fluctuation spectra after corrections by the mode coupling matrix, the beam transfer function, and the map-making transfer function are presented in Figure~\ref{fig:real_final}. As shown, the mosaic images have excess fluctuations, i.e., residual fluctuations over the shot noise, at large angular scales. Errors related with corrections and two systematic errors are also plotted separately along with 1$\sigma$ error of Fourier elements of each bin. Error bars of the filled circles are quadratic sum of them. The effect of shot noise originating from unresolved faint galaxies was estimated by using simulated images of the same size with the mosaic image containing objects fainter than the limiting magnitude. We used the power-law distribution, $N(m) \propto m^{0.23}$ \citep{mai01}, of the number of artificial objects of magnitude, $m$, in unit magnitude interval for the generation of simulated images. The normalization was done based on the number of detected objects in the mosaic image. ARTDATA package of the Image Reduction and Analysis Facility (IRAF; \citealt{tod86}) was used to generate a list of randomly distributed objects and to assign them on the simulation image. Before the estimation of fluctuation spectrum, the mask pattern of the mosaic image was applied to the simulation image. We made 100 simulated images for each band and estimated mean of 100 fluctuation spectra. The overall amplitude of fluctuation is determined by the limiting magnitude. The observed fluctuation spectra at small angular scale well match the shot noise with the limiting magnitude (AB) of 21.4 and 21.8 for 2.4 and 3.2 $\mu$m, respectively.

\section{CONTRIBUTION OF FOREGROUND SOURCES}\label{sec:contribution of foreground sources}

\subsection{Zodiacal Light}\label{sec:zodiacal light}
Because zodiacal light is the most significant contributor to brightness of the near-infrared sky, it is a probable candidate to explain the excess fluctuation of the mosaic image. \citet{pyo12} investigated spatial fluctuation of mid-infrared sky by using the Monitor field data set of \textit{AKARI} and found very small residual fluctuation over photon noise. Since we expect the fluctuation due to zodiacal light does not depend on the wavelength, we adopted their results of $\sim$0.03\% of the sky brightness at 18 $\mu$m band as an upper limit of fluctuations due to zodiacal light in the near-infrared sky at angular scales larger than 100$''$. If we estimate fluctuations by using the results in the mid-infrared sky, the upper limits of fluctuations due to zodiacal light become 0.034 and 0.021 nW m$^{-2}$ sr$^{-1}$ at 2.4 and 3.2 $\mu$m, respectively. Due to the motion of the Earth, IPD column along the line of sight toward the NEP continuously changes during observation period. The fluctuation due to zodiacal light could have been additionally reduced during mosaicking because subsequent pointed observations toward the same field might detect different shape of spatial fluctuations in the zodiacal light. The estimated upper limits due to zodiacal light are clearly smaller than the observed excess fluctuations.

\subsection{Galactic and Extragalactic Components}\label{sec:galactic and extragalactic Components}
Diffuse Galactic light (DGL), starlight scattered by the interstellar dust, can also be a candidate of the excess fluctuation. Because DGL traces the distribution of the interstellar dust, we examined far-infrared 90 $\mu$m band that was taken by Far-Infrared Surveyor\footnote{FIS images are available at http://darts.jaxa.jp/ir/akari} (FIS; \citealt{kawa07}) on board \textit{AKARI}. Slow scan observation with a scan speed of 15 arcseconds per second and a reset interval of 2 seconds were carried out to cover the NEP deep region. We processed the raw pointed data with the official FIS slow-scan toolkit \citep{Verdugo07}. The data reduction tool removed bad data and cosmic rays and corrected for the variation of the detector responsivity. After the dark current subtraction and flat fielding, undetected glitches and other artifacts were removed. The uncertainties of flux calibration is no more than 20\% for the 90 $\mu$m band \citep{kawa07}. The processed images were combined to make a mosaic image using the software Montage.

\begin{figure}
\begin{center}
\includegraphics[scale=1.0, bb=0 0 321 242]{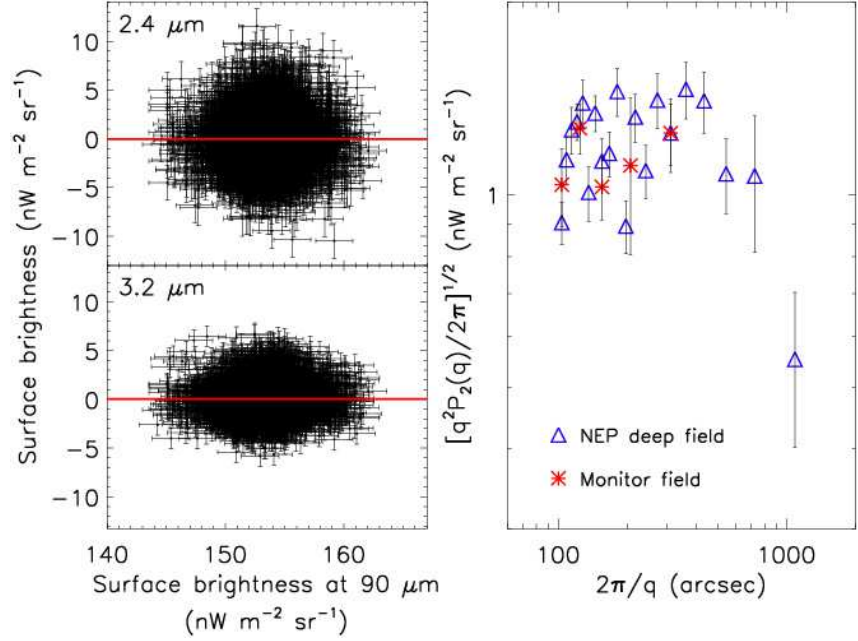}
\caption{Left: correlations between the near-infrared mosaic images and the far-infrared mosaic image at 90 $\mu$m taken by \textit{AKARI}/FIS. Error bars represent the standard deviations of the noise images. Linear fitted lines are shown as solid lines. The slopes of fitted lines are almost zero (10$^{-7}$ or so), indicating that there are no correlations between near- and far-infrared bands. Right: fluctuation spectra of the far-infrared mosaic image at both the Monitor field (asterisk) and the NEP deep field (open triangle). Because shot noise is a dominant component at small angular scale, we focused on fluctuations at angles larger than 100$''$. Error bars show 1$\sigma$ error.}\label{fig:fis}
\end{center}
\end{figure}

To remove the contribution of discrete objects in the far-infrared mosaic image, we masked pixels whose values exceed three times of the standard deviation from mean pixel value, and one layer of pixel around the masked region was additionally masked. About 5\% of pixels were masked in this procedure. To estimate the correlation between images with different pixel resolutions, the angular resolution of the near-infrared mosaic image was adjusted to that of the far-infrared mosaic image, and then a mask of the near-infrared mosaic image was applied on the far-infrared image. The left panel of Figure~\ref{fig:fis} shows that there is no clear correlation between near- and far-infrared images. However, as discussed earlier, because there are some artificial structures in the near-infrared mosaic images, this method has some limited implications. For this reason, we made another examination. The right panel of Figure~\ref{fig:fis} shows fluctuation spectra of the far-infrared image at both the Monitor field and the NEP deep field. We focused on the fluctuations at angular scales larger than 100$''$ where the shot noise becomes small. The far-infrared images of both fields have similar fluctuation powers at angular scales up to 300$''$, and the NEP deep field does not show any significant increase of fluctuation even at larger angles. Such a result represents that the NEP deep field has similar contribution of DGL to the Monitor field. Therefore, we can say that the contribution of DGL is negligible at the NEP deep field by using the result in the right panel of Figure~\ref{fig:fis}.

\begin{figure}
\begin{center}
\includegraphics[scale=0.8]{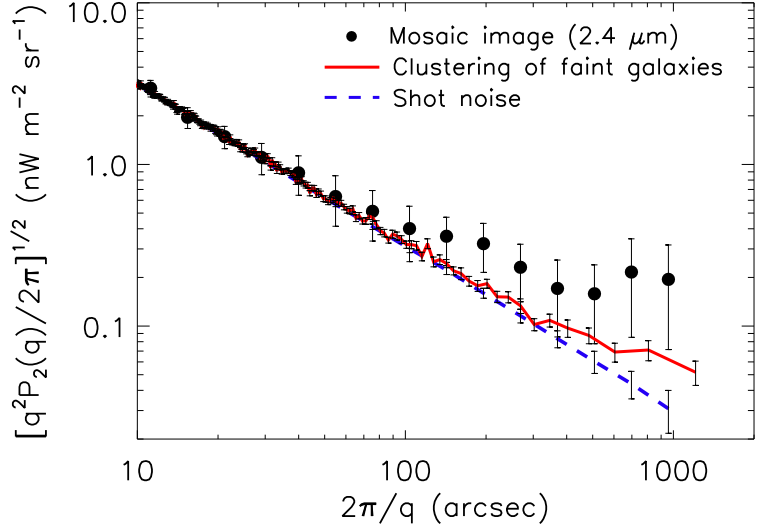}
\caption{Simulation to estimate the clustering of faint objects. Filled circle is fluctuation spectrum of the mosaic image of 2.4 $\mu$m, and dashed line is shot noise of 2.4 $\mu$m. Solid line shows the contribution of the clustering of faint objects, which is estimated by using sources fainter than 21.5 AB mag at $K_{s}$ band of CFHT WIRCam. Mean contribution of the clustering of faint objects, i.e., subtraction of the dashed line from the solid line in quadrature, at angles larger than 300$''$ is 0.059 nW m$^{-2}$ sr$^{-1}$. Error bar of the solid line represents 1$\sigma$ error, and that of the dashed line means a standard deviation of 100 simulations.}\label{fig:cfg}
\end{center}
\end{figure}

The clustering of unresolved galaxies fainter than the limiting magnitude of the mosaic image can also be a candidate for the excess fluctuations. Like the previous study with the Monitor field, the feature of observed excess fluctuation with blue spectrum also shows inconsistency with the study by \citet{cha08} who proposed that a fraction of fluctuations in the near-infrared background could be induced by red dwarf galaxies. We used the deep imaging observations at $K_{s}$ band ($\lambda_{c}$: 2.146 $\mu$m) with Canada-France-Hawaii Telescope (CFHT) Wide-field InfraRed Camera (WIRCam)\footnote{http://www.cfht.hawaii.edu/Instruments/Imaging/WIRCam} for the NEP deep field \citep{oi14} for the estimation of the contribution of the clustering of unresolved faint galaxies to the excess fluctuations. To extract sources in $K_{s}$ band image, we used SExtractor \citep{ber96} with the following parameters: DETECT THRESH=2, DETECT MINAREA=3, BACKSIZE=64, BACK FILTERSIZE=3. Detection limit was estimated by using simulated sources and the same parameter sets of SExtractor to recover them. The 50\% detection limit is about 23.5 AB mag that is deeper than the limiting magnitude of N2 band and is sufficiently sensitive to detect main contribution from the clustering of unresolved faint galaxies. The simulation was performed using the detected objects at $K_{s}$ band in the area (dotted line box in Figure~\ref{fig:imgo}) that covers most regions used in this study. We prepared a simulation image having the same size with the area with dotted line box and assigned detected objects at the same position on the simulation image. When we make a simulated image with objects fainter than 21.5 AB mag at $K_{s}$ band, fluctuation power of the simulation at small angular scale becomes the same as that of N2 band mosaic image or shot noise. A result of the simulation is presented as a solid line in Figure~\ref{fig:cfg}, which shows larger fluctuation than shot noise at large angular scales. We obtained 0.059 nW m$^{-2}$ sr$^{-1}$ as a mean contribution of the clustering of unresolved faint galaxies at angular scales larger than 300$''$, which is 2.8 times smaller than the observed excess fluctuation at the same angular scale. Hence we conclude that the contribution of the unresolved faint low-redshift galaxies is not enough to explain the observed excess fluctuations.

\citet{sull07} also investigated angular power spectra of resolved sources at $J$, $K_{s}$, and $L$ bands. The previous study with the Monitor field \citepalias{mat11} used their model to estimate the contribution of unresolved faint low-redshift galaxies and showed that the estimated contribution is small compared to the observed fluctuation at 2.4 $\mu$m. Recently, \citet{hel12} examined fluctuations in the infrared background from known galaxy populations by using large sets of luminosity functions. They applied their model to actual measurements and concluded that the known galaxy population is not likely to be a candidate to explain the observed fluctuations at arcminute scales. Although it is hard to compare their results with ours quantitatively because our simulation only used the detected objects at $K_{s}$ band without completeness correction at the faint-end of luminosity function, every study predicts that the contribution of unresolved faint low-redshift galaxies to the fluctuation power is small compared to the observational measurements.

\section{DISCUSSION}

\begin{figure}
\begin{center}
\includegraphics[scale=0.8]{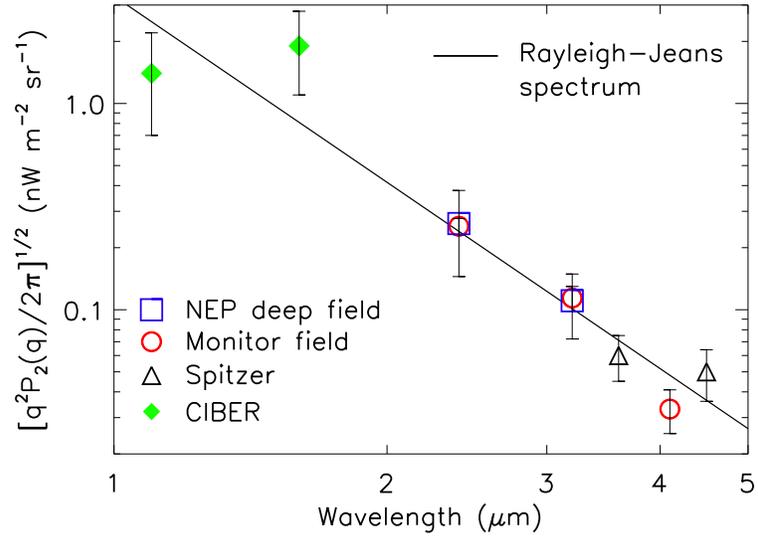}
\caption{Mean excess fluctuations at angles larger than 100$''$. Open square represents the result of the NEP deep field, and open circle is that of the Monitor field \citepalias{mat11}. Open triangle is the result of \textit{Spitzer} \citep{kash05} at a few hundreds arcseconds, and filled diamond is that of CIBER \citep{zem14}. Solid line represents a Rayleigh-Jeans spectrum ($\sim$$\lambda$$^{-3}$), which is fitted by the open circle.}\label{fig:excess}
\end{center}
\end{figure}

By using observational data, we found excess fluctuation in the near-infrared background at angular scales up to 1000$\arcsec$. We estimated a mean of the excess fluctuations of 0.26 and 0.11 nW m$^{-2}$ sr$^{-1}$ for 2.4 and 3.2 $\mu$m bands, respectively, at angular scales larger than 100$''$. In Figure~\ref{fig:excess}, we plotted the results (open squares) and those of the Monitor field (open circles). In addition to the similar level of excess fluctuations, the color of fluctuation components of the NEP deep field is found to be similar to that of the Monitor field: both of them show blue spectra, as first observed by the \textit{IRTS} \citep{mat05} in the excess sky brightness.

\begin{figure}
\begin{center}
\includegraphics[scale=0.8]{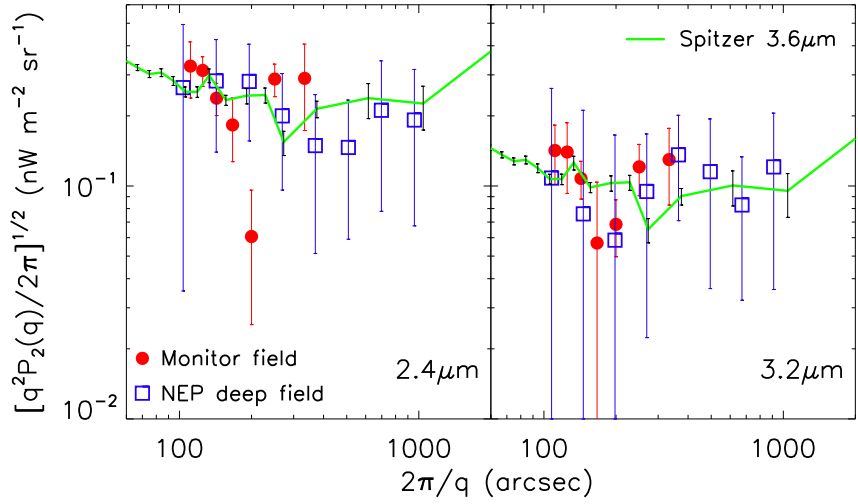}
\caption{Excess fluctuation of the Monitor field (filled circles) and the NEP deep field (open squares). Their error bars are estimated by the error propagation formula with the error of the shot noise. Solid lines represent observational result of \textit{Spitzer} data at 3.6 $\mu$m \citep{kash12}, which is adjusted by using Rayleigh-Jeans spectrum ($\sim$$\lambda$$^{-3}$) for comparison with the result of \textit{AKARI}.}\label{fig:moni_deep}
\end{center}
\end{figure}

By using the result of the NEP deep field, we have extended fluctuation analysis of the Monitor field at angular scales up to 1000$''$. In Figure~\ref{fig:moni_deep}, the result of the Monitor field (filled circles) is combined with that of the NEP deep field (open squares). The fluctuation spectrum of the NEP deep field connects smoothly with that of the Monitor field and extends without showing any significant variation of fluctuation power at angular scales larger than a few hundreds arcseconds. For comparison with the result of \textit{AKARI}, that of \textit{Spitzer} at 3.6 $\mu$m \citep{kash12} is plotted, which is adjusted by the assumption that the spectrum of fluctuating components at the near-infrared wavelength is close to Rayleigh-Jeans spectrum ($\sim$$\lambda$$^{-3}$). It is seen that the average excess fluctuation of \textit{AKARI} is consistent with the result of \textit{Spitzer} at angular scales between 100$''$ and 1000$''$. 

Although the origin of such excess fluctuation is not yet well understood, we could exclude zodiacal light, DGL, and unresolved faint galaxies at low-redshift. One of the remaining possibilities is the contribution from the very early objects at $z\gtrsim$ 10. Theoretical studies have been performed to estimate fluctuation due to the clustering of high-redshift objects. Their results show that fluctuations of high-redshift objects are different from those of low-redshift objects. Our result can be explained from this point of view. Some theoretical studies with biased halo model show the fluctuation powers with a turnover feature at around 10$'$ \citep{coo04,kash04}, while some other studies predict gradually decreasing fluctuation powers toward large angular scales by considering nonlinear biasing of halos \citep{fern10,fern12,coo12a}. The result of this study prefers the latter case, as shown in Figure~\ref{fig:moni_deep}, because our result does not show any significant turnover feature at up to around 10$'$. However, it should be interpreted carefully because there are only a small number of Fourier elements in the last few data points. Meanwhile, although TeV $\gamma$-ray observation prefers lower level of background brightness than observational measurements in the near-infrared bands to explain absorption feature in $\gamma$-ray spectra of blazars \citep{aha06,gil12,Meyer12,hess13}, it is not a stringent constraint because alternative explanation of $\gamma$-ray spectra of blazars is possible using secondary TeV $\gamma$-ray photons produced by interaction between cosmic rays of blazars and EBL \citep{ess10,essku10}.

However, theoretical studies predicted smaller fluctuations than observations by about an order of magnitude. Recently, \citet{yue13a} investigated the contribution of high-redshift galaxies at $z >$ 5 to the near-infrared background by using simulations with an analytical model. They predicted that fluctuation amplitude due to $z >$ 5 galaxies would be less than 0.01 nW m$^{-2}$ sr$^{-1}$ at a few hundreds arcseconds in the wavelength range 1.0$\sim$4.5 $\mu$m, which is 10 times smaller than the excess fluctuations in this study and recent observational studies using \textit{Spitzer} imaging data \citep{coo12b,kash12}. Therefore, further studies for both theory and observation are required to identify the origin of excess fluctuation in the near-infrared background.

\citet{coo12b} suggested that the extended emission due to IHL can be a candidate to explain observed fluctuation in the near-infrared background. Although it is an interesting idea, it may be challenging to verify observationally. As discussed in their work, simple masking is not an efficient method to remove diffuse components of faint galaxies. To identify extended components around faint galaxies and to subtract them accurately, we need imaging data with high sensitivity and a sophisticated method for modeling them.

Recently, the rocket-borne instrument, CIBER, measured the fluctuation of the sky brightness up to 1$^{\circ}$ angular scale at 1.1 and 1.6 $\mu$m bands \citep{zem14}. Since there exist fluctuations at shorter wavelength than 2 $\mu$m, IHL model is preferred than the contribution from Population III stars in the early universe. They also found that the result of CIBER shows good correlation with the result of \textit{Spitzer} and fluctuation at 1.1 $\mu$m shows some departure from the Rayleigh-Jeans spectrum.

\citet{yue13b} suggested a contribution of direct collapse black holes (DCBHs) at $z\gtrsim$ 12 as a possible candidate for near-infrared fluctuations. They claimed that the DCBH model can explain observed fluctuations at 3.6 and 4.5 $\mu$m and observed cosmic X-ray background/near-infrared background cross-correlation. However, DCBH model is inconsistent with substantial fluctuations even around 1 $\mu$m observed by the CIBER \citep{zem14}.

To investigate various aspects of the excess fluctuation, more observational studies are needed by using individual imaging data with various FOV, angular scales, and wavelength bands. As one of the scheduled programs for the study of the near-infrared background, the Multipurpose Infrared Imaging System (MIRIS; \citealt{Han10,Han14}), launched in Nov. 2013, is observing a large area of 10$^{\circ}$ $\times$ 10$^{\circ}$ centered at NEP for the study of the sky brightness and its fluctuations at 1.1 and 1.6 $\mu$m bands. It will provide new information about the sky brightness in the near-infrared bands at angles up to a few degrees and improve our understanding of the origin of the cosmic infrared background.

\section{SUMMARY}

We studied spatial fluctuations of the sky brightness using the NEP deep survey data obtained with \textit{AKARI} IRC at two wavelength bands, 2.4 and 3.2 $\mu$m. Compared with the Monitor field of \textit{AKARI} used in our previous study, the NEP deep field has larger areal coverage, enabling us to extend fluctuation analysis of the sky brightness at larger angular scales up to 1000$''$. We produced mosaic images for two bands and estimated spatial fluctuations using power spectrum analysis. We found that there is a residual fluctuation over the estimated shot noise even at larger angles than the angular scale of the Monitor field, which can hardly be explained by any known foreground components.

By using excess fluctuation of the NEP deep field, we confirm a blue stellar-like spectrum of excess fluctuation that was found in the previous study with the Monitor field and find the blue spectrum is still recognized even at larger angular scales. The excess fluctuation spectra of the NEP deep field extends smoothly that of the Monitor field at angular scales of a few hundreds arcseconds, and combination of the two results does not show any significant variation of fluctuation power at angular scales between a few hundreds and $\sim$1000$''$. We show that the result of this study is consistent with the study of \textit{Spitzer} at 3.6 $\mu$m. Such results could provide useful constraints for the study of background at the near-infrared wavelength bands.

\acknowledgments

This work is based on observations with \textit{AKARI}, a JAXA project with the participation of ESA. We thank the \textit{AKARI} IRC team for their encouragement and helpful discussions. H.M.L. and M.G.L. were supported by NRF grant No. 2012R1A4A1028713. This research made use of Montage, funded by the National Aeronautics and Space Administration's Earth Science Technology Office, Computation Technologies Project, under Cooperative Agreement Number NCC5-626 between NASA and the California Institute of Technology. Montage is maintained by the NASA/IPAC Infrared Science Archive.

{\it Facility:} \facility{\textit{Akari} (IRC)}.

\clearpage

\clearpage

\begin{deluxetable}{crrr}
\tablecaption{Sinusoidal fitting of zodiacal light\label{tb:fitting}}
\tablewidth{0pt}
\tablehead{\colhead{Band} & \colhead{\textit{a}} & \colhead{\textit{b}} & \colhead{\textit{c}} \\
\colhead{} & \colhead{(nW m$^{-2}$ sr$^{-1}$)} & \colhead{(deg.)} & \colhead{(nW m$^{-2}$ sr$^{-1}$)}}
\startdata
N2 & 11.70 $\pm$ 0.010 & 66.20 $\pm$ 0.089 & 113.92 $\pm$ 0.016 \\
N3 & 6.70 $\pm$ 0.004 & 61.62 $\pm$ 0.055 & 71.26 $\pm$ 0.005 \\
\enddata
\end{deluxetable}

\clearpage

\begin{deluxetable}{lcc}
\tablecaption{Basic parameters of the mosaic image of the NEP deep field\label{tb:data_deep}}
\tablewidth{0pt}
\tablehead{\colhead{Item} & \colhead{2.4 $\mu{m}$} & \colhead{3.2 $\mu{m}$}}
\startdata
R.A.(J2000) DEC.(J2000) & 17$^{h}$55$^{m}$24$^{s}$ & 66$^{\circ}$37$'$32$''$ \\
Number of pointed observations & 17 & 17 \\
Number of images & 75 & 68 \\
Mean observation time (minutes) & 12.0 & 10.9 \\
Mean number of stacked images & 11 & 10 \\
Limiting magnitude (AB mag) & 21.4 & 21.8 \\
Percentage of remaining pixels (\%) & 39.6 & 39.5 \\
Average sky brightness (nW m$^{-2}$ sr$^{-1}$) & 113.92 $\pm$ 0.016 & 71.26 $\pm$ 0.005 \\
\enddata
\end{deluxetable}

\end{document}